\documentclass{article}
\usepackage{spconf_arxiv}
\usepackage{amsmath,graphicx}
\usepackage{array, amsfonts}
\usepackage{booktabs}
\usepackage{multirow}
\usepackage{makecell}
\usepackage{color}
\usepackage{bbding}
 
\title{Parameter-efficient transfer learning of Pre-trained \\
Transformer models for speaker verification using adapters}
\name{Junyi Peng$^{1}$, Themos Stafylakis$^{2}$, Rongzhi Gu$^{3}$, Old\v{r}ich Plchot$^{1}$, Ladislav Mo\v{s}ner$^{1}$ \thanks{The work was supported by Czech National Science Foundation (GACR) project NEUREM3 No. 19-26934X, Czech Ministry of Interior project No. VJ01010108 "ROZKAZ", Czech Ministry of Education, Youth and Sports from project no. LTAIN19087 "Multi-linguality in speech technologies", Horizon 2020 Marie Sklodowska-Curie grant ESPERANTO, No. 101007666, and by Tencent AI Lab Rhino-Bird Gift Fund. Computing on IT4I supercomputer was supported the Ministry of Education, Youth and Sports of the Czech Republic through the e-INFRA CZ (ID:90140).
}}

\secondname{Luká\v{s} Burget$^{1}$, Jan \v{C}ernocký$^{1}$}

\address{
$^1$Brno University of Technology, Faculty of Information Technology, Speech@FIT, Czechia \\
$^2$Omilia - Conversational Intelligence, Athens, Greece\\
$^3$Tencent AI Lab, Shenzhen, China \\
}
\begin{document}
\ninept

\maketitle

\begin{abstract}
Recently, the pre-trained Transformer models have received a rising interest in the field of speech processing thanks to their great success in various downstream tasks. However, most fine-tuning approaches update all the parameters of the pre-trained model, which becomes prohibitive as the model size grows and sometimes results in overfitting on small datasets. In this paper, we conduct a comprehensive analysis of applying parameter-efficient transfer learning (PETL) methods to reduce the required learnable parameters for adapting to speaker verification tasks. Specifically, during the fine-tuning process, the pre-trained models are frozen, and only lightweight modules inserted in each Transformer block are trainable (a method known as \emph{adapters}). Moreover, to boost the performance in a cross-language low-resource scenario, the Transformer model is further tuned on a large intermediate dataset before directly fine-tuning it on a small dataset.  With updating fewer than 4\% of parameters, (our proposed) PETL-based methods achieve comparable performances with full fine-tuning methods (Vox1-O: 0.55\%, Vox1-E: 0.82\%, Vox1-H:1.73\%).

\end{abstract}
\begin{keywords}
Speaker verification, pre-trained model, adapter, fine-tuning, transfer learning \end{keywords}

\section{Introduction}
\label{sec:intro}
A typical state-of-the-art speaker verification (SV) system is based on comparison of speaker embeddings which are extracted using a deep neural model \cite{snyder2018x,desplanques2020ecapa,zhou2021resnext} trained from scratch on a large-scale speaker-labeled dataset such as Voxceleb \cite{nagrani2017voxceleb,chung2018voxceleb2}. The size (typically more than 10 million parameters) and the architecture based on a series of convolutional layers make it difficult to properly train these extractors in a data-restricted scenario of some low-resource domain (i.e., completely new channel, language, or their combination).

Recently, large pre-trained Transformer models, including Wav2Vec \cite{ baevski2020Wav2Vec}, HuBERT \cite{hsu2021hubert}, WavLM \cite{chen2021wavlm}, and their variants \cite{chen2022unispeech} have significantly boosted the performance in the field of speech processing. The most common way to adapt those general-purpose models to downstream tasks is to fine-tune the whole pre-trained model with a task-oriented back-end (\emph{full fine-tuning}). In \cite{chen2022large}, a strong performance on the speaker verification task was achieved with the ECAPA-TDNN back-end, of which the frame-by-frame input to the back-end was calculated as a weighted combination of outputs of the individual layers of a pre-trained Transformer model. In \cite{peng2022attention}, to shorten the training time, a more lightweight back-end is employed, which consists of an attention layer and a linear layer to extract speaker representations. However, potential shortcomings of such full fine-tuning are the necessity of updating a vast amount of parameters and storing a separate task-related copy of the fine-tuned model parameters for each downstream task or its domain-specific version. This issue will become increasingly problematic with the number of parameters of the pre-trained model growing from hundreds of millions to billions. For example Whisper \cite{radford2022robust} contains 1,550 M parameters.

\begin{figure}[t]
  \centering
  \includegraphics[width=\linewidth]{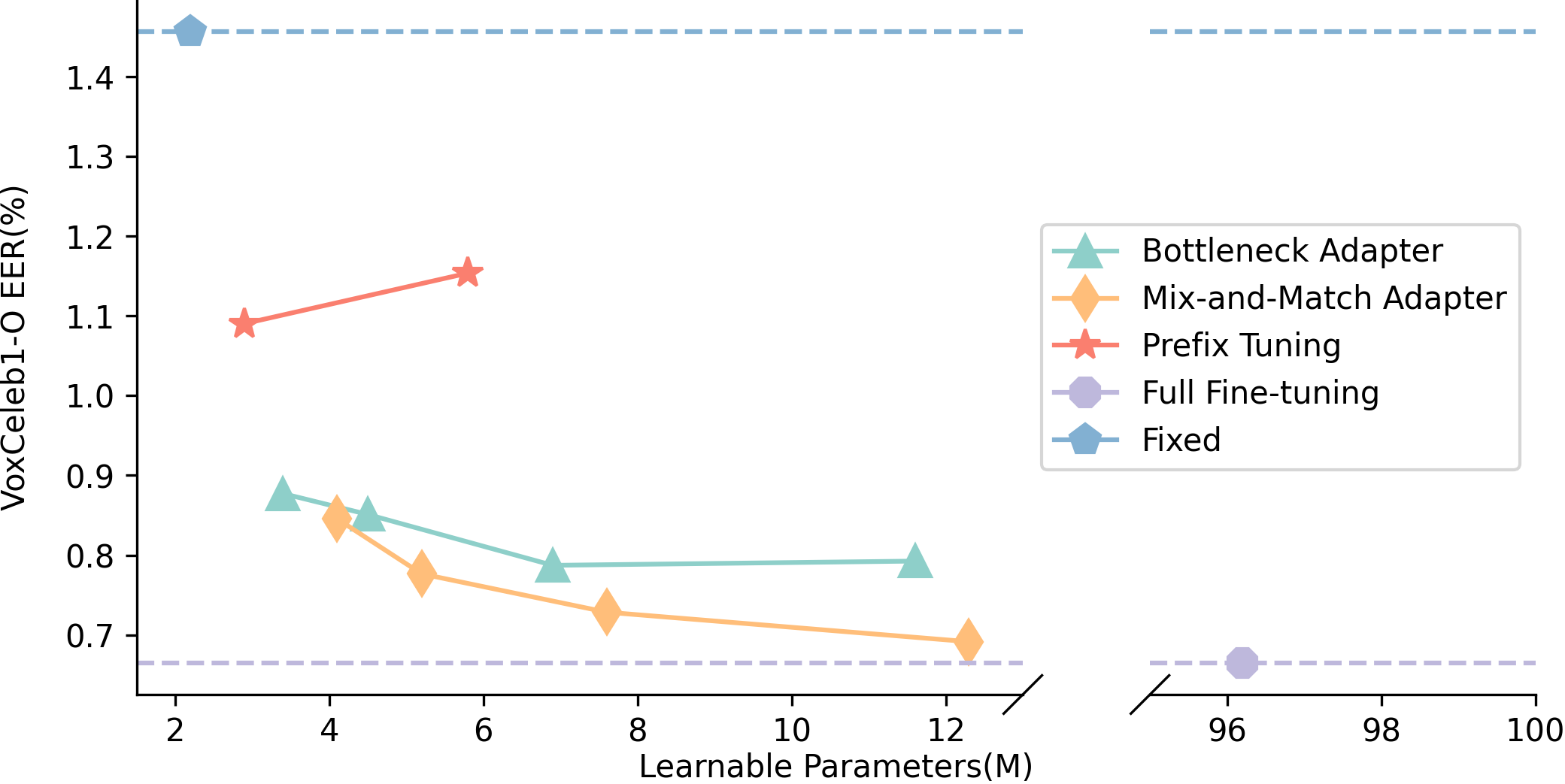}
\caption{Performance change of several PETL approaches on VoxCeleb1-O, when the number of learnable parameters is increased. The learnable parameters include the speaker extractor back-end with constant 2.2M parameters and the PETL module.}
\label{fig:fig1}
\end{figure}

\begin{figure*}[t]
\begin{minipage}[t]{.24\linewidth}
  \centering
  \centerline{\includegraphics[height=5.2cm]{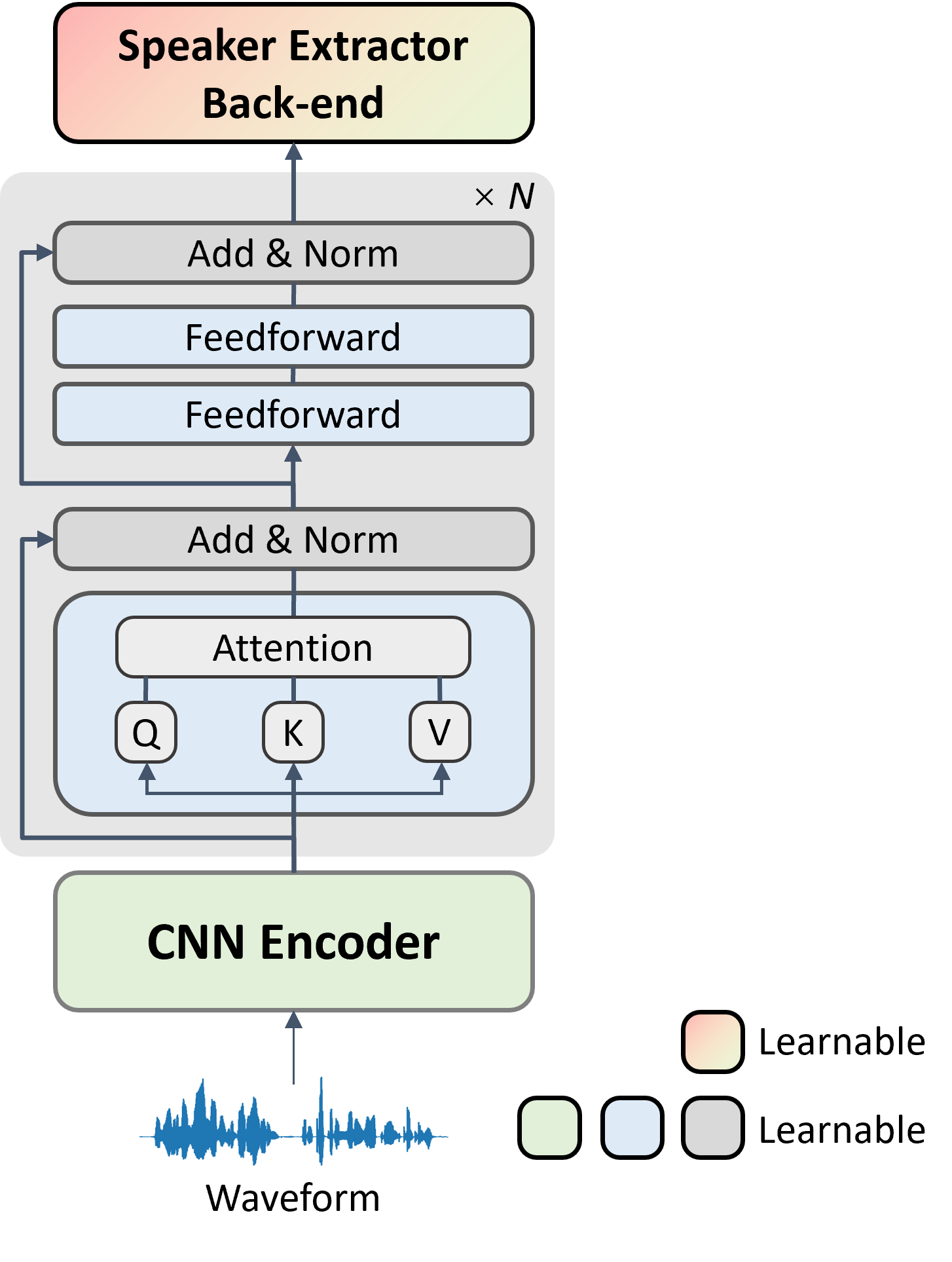}}
  \centerline{(a) Full Fine-tuning}\medskip
\end{minipage}
\begin{minipage}[t]{.24\linewidth}
  \centering
  \centerline{\includegraphics[height=5.2cm]{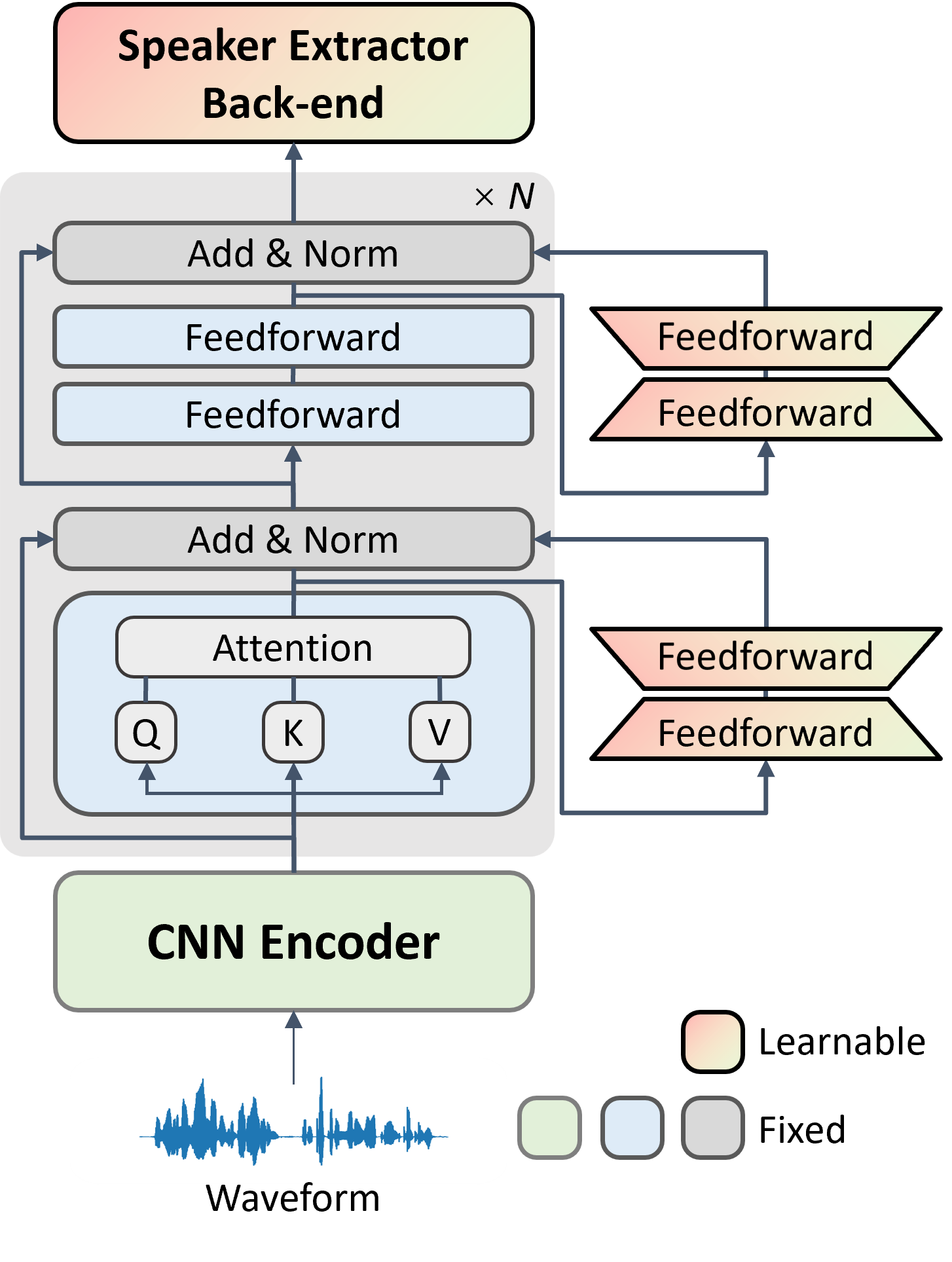}}
  \centerline{(b) Bottleneck Adapter}\medskip
\end{minipage}
\hfill
\begin{minipage}[t]{0.24\linewidth}
  \centering
  \centerline{\includegraphics[height=5.2cm]{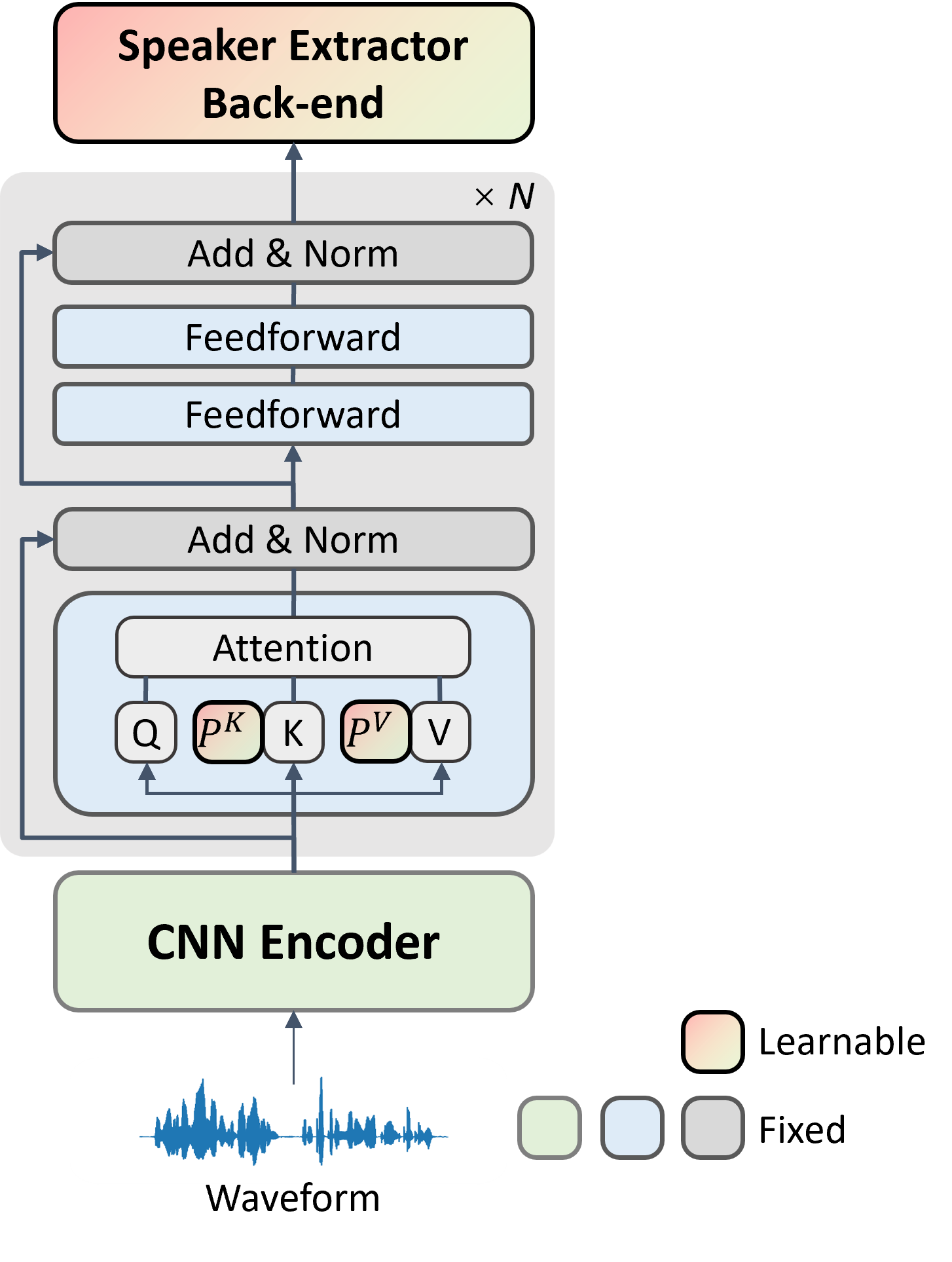}}
  \centerline{(c) Prefix Tuning}\medskip
\end{minipage}
 \hfill
\begin{minipage}[t]{0.24\linewidth}
  \centering
  \centerline{\includegraphics[height=5.2cm]{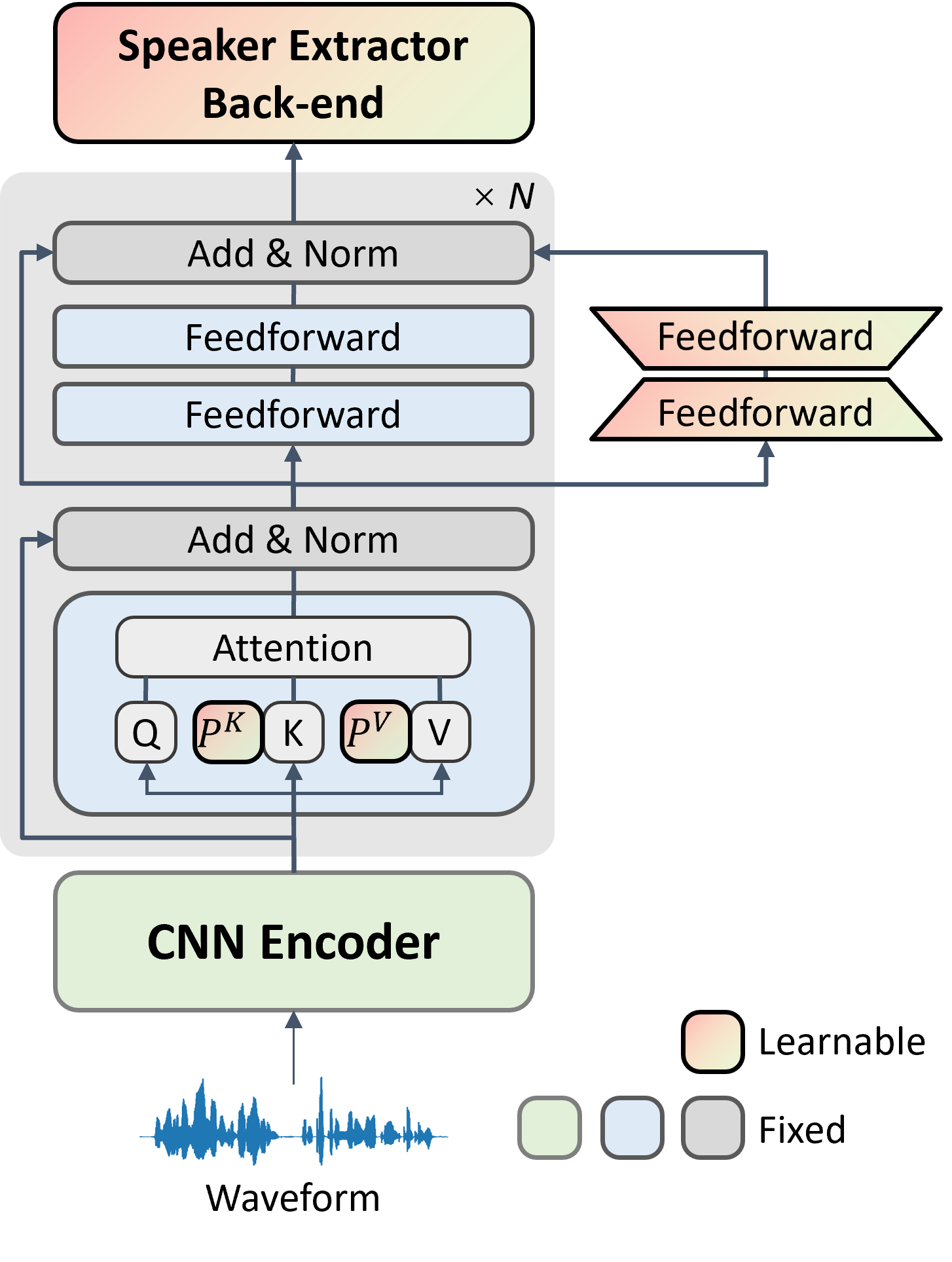}}
  \centerline{(d) Mix-and-Match Adapter}\medskip
\end{minipage} 
  
  \vfill
\caption{Architecture of the pre-trained model and state-of-the-art parameter-efficient methods. For (b)(c)(d), only the inserted lightweight modules and back-end are learnable during fine-tuning, while the pre-trained model is frozen. “Speaker Extractor Back-end” consists of a multi-head factorized attentive pooling (MHFA) and a linear layer to extract speaker representations \cite{peng2022attention}. }
\label{fig:sys}
\end{figure*}

To alleviate this issue, many recent studies have focused on parameter-efficient transfer learning, known as adapter, where additional lightweight modules with task-specific trainable parameters are inserted into the pre-trained model while keeping the entire pre-trained model frozen. For example, in \cite{thomas2022efficient}, a bottleneck adapter \cite{houlsby2019parameter} is applied to the Wav2Vec model and the adapter-based model achieved comparable performance to full fine-tuning by only updating 10\% of the model parameters in ASR tasks. In \cite{le2021lightweight}, pre-trained models are connected with a multilingual denoising auto-encoder for speech-to-text translation through adapter modules. In addition, a more challenging and not extensively explored problem with pre-trained speech models is their adaptation to a low-resource scenario with few trainable parameters. Indeed, most pre-trained models are optimized on English corpora (e.g. LibriSpeech \cite{panayotov2015librispeech}). Such models are supposed to be eminently suitable for downstream tasks transfer learning using the same language. However, there is no reason to believe those pre-trained models can also provide a proper initialization for unseen languages since the distribution of acoustic units might be completely different \cite{zhou2018comparison}. 
When fully fine-tuning on a different dataset, the training process might degenerate the model, resulting in catastrophic forgetting of what was learned during the pre-training phase \cite{mccloskey1989catastrophe}.

To mitigate this issue, in this paper, we first analyze the performance of three different PETL methods, including bottleneck adapter~\cite{houlsby2019parameter}, prefix tuning \cite{li2021prefix}, and mix-and-match adapter \cite{he2021towards}, to transfer the pre-trained model to downstream speaker verification tasks. Then, we explore using model tuned on an intermediate dataset before fine-tuning it to a small out-of-domain (cross-language in our case) dataset. This approach reduces the variance between target and source domains, and improves the robustness and discrimination of learned speaker representations, resulting in boosting the performance in the low-resource setting.
The contributions of our work are as follows:
\begin{itemize}
    \item We demonstrate that the PETL methods can be utilized to effectively adapt large-scale pre-trained transformer models to a specific downstream task (e.g., speaker verification) with few learnable parameters, as shown in Fig \ref{fig:fig1}.
    \item To further boost the performance in the cross-language low-resource scenario, we tune the pre-trained model using an intermediate dataset before fine-tuning it on a small dataset. This achieves state-of-the-art results on the CNCeleb dataset.
    \item Extensive experiments on VoxCeleb corpus \cite{nagrani2017voxceleb,chung2018voxceleb2} show that adapter-based fine-tuning can achieve comparable performance to full fine-tuning through updating less than 4\% of the original model parameters.\footnote{The code will be available with the submission of the final paper.} 
\end{itemize}


\section{Parameter-efficient transfer learning}
\label{sec:PETL}
In this section, we will introduce three state-of-the-art parameter-efficient transfer learning methods, as shown in Fig \ref{fig:sys}. Unless otherwise emphasized, the parameters of pre-trained models are frozen during the fine-tuning process, while only the parameters of lightweight additional modules are trainable.

\noindent\textbf{Bottleneck Adapter:} 
As illustrated in Fig \ref{fig:sys} (b), bottleneck adapter \cite{houlsby2019parameter} is inserted into each Transformer block of a pre-trained model after multi-head attention layers and feed-forward layers with a residual connection. The bottleneck adapter layer consists of a down projection layer $\mathbf{W}_\textit{down} \in \mathbb{R}^{D_\textit{hidden}\times D_\textit{bottleneck}}$, a up projection layer $\mathbf{W}_\textit{up} \in \mathbb{R}^{D_\textit{bottleneck}\times D_\textit{hidden}}$, as well as a nonlinear activation function $f(\cdot)$. Its frame-by-frame output
\begin{equation}
\label{b-adapter}
\begin{split}
\mathbf{H}_\textit{out} = \mathbf{H}_\textit{in} + \mathbf{W}_\textit{up}f(\mathbf{W}_\textit{down}\mathbf{H}_\textit{in})
\end{split}
\end{equation}
is of the same size as the input $\mathbf{H}_\textit{in} \in \mathbb{R}^{ T\times D_\textit{hidden}}$, where $T$ is the length of the input sequence.

\noindent\textbf{Prefix Tuning:}
Different from the bottleneck adapter that utilizes the outputs of different layers, prefix tuning \cite{li2021prefix} adds a set of $l$ learnable vectors (virtual tokens) as additional keys and values to each head of the multi-head attention module in each Transformer block. During the fine-tuning phase, these learnable vectors are expected to capture task-related information and adapt the pre-trained model to the downstream task, as shown in Fig \ref{fig:sys} (c). We denote the linear projections of \textit{queries}, \textit{keys} and \textit{value} of each head of each attention module\footnote{We omit the layer and head indices in the symbols to keep the notation uncluttered.} as $\mathbf{W}_Q$, $\mathbf{W}_K$ and $\mathbf{W}_V\in \mathbb{R}^{D_\textit{hidden} \times D_\textit{proj}}$, respectively. $D_\textit{proj}$ is the projected dimension of each head. Typically, $D_\textit{proj}=D_\textit{hidden}/H$, where $H$ is the total number of heads.
The attention maps are evaluated as:
\begin{equation}
\label{prefix}
\begin{split}
Attn(Q,K_\textit{prefix},V_\textit{prefix}) & = softmax\left(\frac{\mathbf{Q}\mathbf{K}^T_\textit{prefix}}{\sqrt{D_\textit{proj}}}\right)\mathbf{V}_\textit{prefix} \\
\mathbf{K}_\textit{prefix} &= concat(\mathbf{P}_{K},\mathbf{W}_K \mathbf{H}_\textit{in}) \\
\mathbf{V}_\textit{prefix} &= concat(\mathbf{P}_{V},\mathbf{W}_V \mathbf{H}_\textit{in}) \\
\end{split}
\end{equation}
where two learnable matrices (the two sets of virtual tokens) $\mathbf{P}_K$, $\mathbf{P}_V \in \mathbb{R}^{l \times D_\textit{proj}}$ are prepended to the original \textit{keys} and \textit{values}, respectively.


\noindent\textbf{Mix-And-Match Adapter:} To combine and unify the two aforementioned PETL methods, in \cite{he2021towards}, a new variant, named mix-and-match adapter, was applied. As illustrated in Fig \ref{fig:sys} (d), the MAM uses an adapter block processing the hidden representation parallel to the feedforward block. Additionally, it leverages a small prefix tuning module to generate task-related attention maps.
\vspace{-1.5mm}

\section{Fine-tuning via intermediate dataset}
HuBERT-style models consume masked frame-level features to predict a pre-determined discrete target during the unsupervised pre-training phase. When using those models to deal with a small dataset, the pre-learned parameters are supposed to be ideally appropriate for the downstream tasks. However, when fully fine-tuning to a cross-language low-resource scenario, the learning process often gets stuck in local minima. This might be because the target dataset may have a different distribution of acoustic units that is unseen during pre-training. Thus, to improve the robustness, in this paper, we tune the pre-trained model on a large intermediate supervised SV dataset before fine-tuning it to a small dataset. With this two-step tuning scheme, the task-related model is expected to be reasonably close to the proper setting for the low-resource target task.
\vspace{-0.5mm}

\section{Experiments}
\label{sec:exp}
\vspace{-3mm}
\subsection{Setup}
\textbf{Data-sets:} The SV performance is evaluated on the VoxCeleb \cite{nagrani2017voxceleb,chung2018voxceleb2} and CNCeleb \cite{fan2020cn,li2022cn} corpora, both are widely used text-independent speaker verification datasets. For VoxCeleb, the training set is the development set of VoxCeleb2. The performance is evaluated on \textit{VoxCeleb1-O}, \textit{VoxCeleb1-E}, and \textit{VoxCeleb1-H} trials. For CNCeleb, the model is fine-tuned on three different training dataset, namely \textit{CNCeleb1-S1}, \textit{CNCeleb1-S2}, \textit{CNCeleb1} and \textit{CNCeleb.T}, containing 200, 400, 800 and 2800 speakers, respectively. CNCeleb.T is a combination of CNCeleb1-dev and CNCeleb2. The evaluation part \textit{CNCeleb-E} contains 18,849 utterances from 200 speakers. Besides, all training datasets are augmented by adding noise and reverberation.  

\noindent\textbf{Implementation details:}
In this work, we utilize two types of pre-trained models: 1) The Base models, including WavLM Base+ and HuBERT Base, contain a CNN encoder and 12 layers of Transformer. The dimension of the Transformer output $D_{hidden}$ is 768. The total number of parameters of those models is around 94M; 2) The Large model has 24 transformer blocks with 1024-dimensional output resulting in 316M parameters. All experiments are conducted on 8 A100 GPUs with 10 epochs optimizing AAM-softmax \cite{deng2019arcface} with a margin of 0.2 and scaling of 30. To speed up the training, the learning rate is decreased by 5\% each epoch. The duration of input raw waveforms is set to 3 seconds. The mini-batch size of 120 is chosen for training models. We also adopt large margin fine-tuning (LM-FT) \cite{thienpondt2021idlab} to further boost performance. Specifically, we input longer (5 seconds) waveforms and set the margin to 0.5 for additional 2 tuning epochs. 

\noindent\textbf{Performance Metrics:}
Both equal error rate (EER) and minimum detection cost function (minDCF) are employed to measure the performances of speaker verification systems. The prior target probability $P_{\textit{tar}}$ is set to 0.01 or 0.05, for DCF1 and DCF5, respectively. $C_{\textit{fa}}$ and $C_{\textit{miss}}$ are set to 1.0.

\begin{table}[t] 
  \caption{Results on the VoxCeleb1-O dataset. For a fair comparison, all methods use WavLM Base+ as the frozen pre-trained model. }
  \label{tab:1}
  \centering
    \scalebox{0.9}{
    \begin{tabular}{l|c|l|c}  
    \hline  
    Adapter & Params & Dim & EER (\%) \\
    \hline   
    \hline
    \multirow{3}*{Bottleneck Adapter} 
    & 4.7M & $D_{bottleneck}=128$ & 0.78 \\
    & 2.3M & $D_{bottleneck}=64$ & 0.85\\
    & 1.2M & $D_{bottleneck}=32$ &0.87 \\
        \hline
        \multirow{2}*{Prefix tuning} & 3.6M & $l=200$ & 1.15 \\
    & 0.7M & $l=40$ & 1.09 \\
    \hline 
        \multirow{3}*{MAM Adapter} 
    & 5.4M & $D_\textit{bottleneck}=256, l=40$ & \textbf{0.72} \\
    & 3.0M & $D_\textit{bottleneck}=128, l=40$ & 0.77\\
    & 1.9M & $D_\textit{bottleneck}=64, l=40$ & 0.84 \\

    \hline

    \end{tabular}
    }
    \vspace{-0.5cm}
\end{table}

\begin{table*}[t]
  \caption{Results on the Voxceleb1 dataset and extended test sets. All models are trained on VoxCeleb2-dev, except \dag – its training data consists of Vox2-dev and Vox1-dev. LM-FT denotes large-margin fine-tuning. The learnable back-end MHFA \cite{peng2022attention} contains 2.2M parameters. }
  \label{tab:SOTA}
  \centering
  \scalebox{0.9}{
    \begin{tabular}{l|c|c|c|c|c|c|c|c|c|c}  
    \hline  
    \multicolumn{1}{c|}{\multirow{2}{*}{Front-end Model}}&\multicolumn{1}{c|}{\multirow{2}{*}{Params }}&\multicolumn{3}{c|}{VoxCeleb1-O}&\multicolumn{3}{c}{VoxCeleb1-E}&\multicolumn{3}{|c}{VoxCeleb1-H}\cr\cline{3-11}  & &EER(\%)&DCF1 &DCF5&EER(\%)&DCF1&DCF5 &EER(\%)& DCF1&DCF5 \\
    \hline    
    \hline
    ECAPA-TDNN \cite{kwon2021ins} & 14.7M & 0.90 & - & 0.081 & 1.11 & - & 0.077 & 2.32 & - & 0.155\\
    wav2vec-TDNN $^{\dag}$ \cite{novoselov2022robust} & 317M + 3M & 0.84 & 0.058 & - & - & - & - & - & - & - \\
    UnispeechSAR\_BASE-TDNN \cite{chen2021wavlm} & 94M+6M & 1.00 & - & -  & 0.93 & -  & - & 1.87 & -  & - \\
    \hline
    \multicolumn{8}{l}{Pre-trained Model: \textbf{HuBERT BASE}, Back-end: \textbf{MHFA}} \\
    \hline
    Full fine-tuning & 94.6M+2.2M & 0.82 & 0.114 & 0.061 & 1.13 & 0.122 & 0.073 & 2.43 & 0.244 & 0.014 \\
    Fixed  & 0.0M + 2.2M &  1.96 & 0.221 & 0.525 & 2.27 & 0.252 & 0.152 & 4.62 & 0.416 & 0.131 \\
   Bottleneck Adapter & 4.7M + 2.2M & 0.98 & 0.138 & 0.068 & 1.21 & 0.137 & 0.081 & 2.61 & 0.260 & 0.162 \\
   Prefix Tuning & 3.6M + 2.2M & 1.55 & 0.193 & 0.107 & 1.74 & 0.198 & 0.118 & 3.86 & 0.356 & 0.233\\
   MAM Adapter & 5.4M + 2.2M & 0.96 & 0.130 & 0.065 & 1.18 & 0.133 & 0.079 & 2.56 & 0.261 & 0.161 \\
   \hline
       \multicolumn{8}{l}{Pre-trained Model: \textbf{WavLM BASE+}, Back-end: \textbf{MHFA}} \\
    \hline
    Full fine-tuning & 94.7M+2.2M & 0.66 & 0.074 & 0.045 & 0.89 & 0.097 & 0.056 & 1.90 & 0.190 & 0.119 \\
    Full fine-tuning [LM-FT] \cite{peng2022attention} & 94.7M+2.2M & 0.59 & 0.069 & 0.041 & 0.79 & 0.089 & 0.050 & 1.73 & 0.177 & 0.107 \\
    Fixed  & 0.0M + 2.2M & 1.45 & 0.167 & 0.098 & 1.64 & 0.191 & 0.111 & 3.45 & 0.330 & 0.207 \\
   Bottleneck Adapter & 4.7M + 2.2M & 0.78 & 0.073 & 0.052 & 0.96 & 0.108 & 0.063 & 2.10 & 0.215 & 0.131 \\
   Prefix Tuning & 3.6M + 2.2M & 1.15 & 0.128 & 0.068 & 1.27 & 0.145 & 0.083 & 2.69 & 0.253 & 0.161 \\
   MAM Adapter & 5.4M + 2.2M & 0.72 & 0.086 & 0.052 & 0.92 & 0.107 & 0.059 & 2.05 & 0.212 & 0.132 \\
   MAM Adapter [LM-FT] & 5.4M + 2.2M & 0.61 & 0.058 & 0.041 & 0.88 & 0.099 & 0.055 & 1.90 & 0.193 & 0.119 \\
   \hline
    \multicolumn{8}{l}{Pre-trained Model: \textbf{WavLM Large}, Back-end: \textbf{MHFA}} \\
    \hline
    Full fine-tuning [LM-FT] & 316M + 2.2M & 0.49 & 0.081 & 0.041 & 0.70 & 0.091 & 0.051 & 1.70 & 0.177 & 0.105 \\
   MAM Adapter [LM-FT] & 12.5M + 2.2M & 0.55 & 0.065 & 0.038 & 0.82 & 0.091 & 0.050 & 1.73 & 0.166 & 0.104 \\

    \hline
    \end{tabular}
    }
    \vspace{-0.5cm}
\end{table*}

  



\begin{table}[t]  
  \caption{Results on the CNCeleb-E dataset with different size of training dataset. \textit{CNCeleb1-S1 and -S2} denotes a subset of CNCeleb1 with randomly selected 200 and 400 speakers, respectively. [Int. D] means Intermediate dataset, i.e. the Transformer model is continually tuned on Vox2-dev before fine-tuning. WavLM Base+ is used as pretrained model.}
  
  \label{tab:3}
  \centering
    \scalebox{0.65}{
    \begin{tabular}{l|c|c|c|c|c|c|c|c}  
    \hline  
    \multicolumn{1}{c|}{\multirow{3}{*}{Training Dataset}}&\multicolumn{2}{c|}{\multirow{1}{*}{\makecell{CNCeleb1-S1}}}&\multicolumn{2}{c|}{\multirow{1}{*}{\makecell{CNCeleb1-S2}}}&\multicolumn{2}{c|}{\multirow{1}{*}{\makecell{CNCeleb1}}}&\multicolumn{2}{c}{\multirow{1}{*}{\makecell{CNCeleb.T}}}\cr & \multicolumn{2}{c|}{\# 200 Spk} & \multicolumn{2}{c|}{\# 400 Spk} &  \multicolumn{2}{c|}{\# 800 Spk} & \multicolumn{2}{c}{\# 2800 Spk} \cr \cline{2-9} & EER & DCF1 & EER & DCF1 & EER & DCF1 & EER & DCF1 \\
    \hline   
    \hline 
    Sparse FilterBank \cite{peng2022learnable} & - & - & - & - & 12.25 & 0.5391 & - & - \\
    Modified x-vector \cite{li2022real}  & - & - & - & - & 11.05 & - & - & - \\
    R-vector \cite{chen2021self} & - & - & - & - & 8.86 & - & -  & - \\
    ECAPA-TDNN \cite{zeng2022attention} & - & - & - & - & - & - & 8.93 & 0.5043 \\
    \hline
    FT & 17.43&0.8624 & 12.25 & 0.6384 & 10.05&0.5000 & 7.93&0.4079 \\
    FT [Int. D] & 9.25&0.5053 & 8.83 & \textbf{0.4515} & 8.45&0.4145 & 7.71&0.4057 \\
    MAM Adapter & 13.73&0.7029 & 11.19 & 0.5715 & 9.45& 0.4745 & 7.52 & 0.4072 \\
    MAM Adapter [Int. D] & \textbf{9.12}&\textbf{0.4963} & \textbf{8.58} & 0.4671 &  \textbf{7.94}&\textbf{0.4087} & \textbf{6.89}&\textbf{0.3784} \\

    \hline 

    \end{tabular}

}
\vspace{-0.3cm}
\end{table}

\subsection{Analysis of PETL methods}
We first investigate the performance of the three PETL variants. In the field of natural language processing, \emph{prefix tuning} attains comparable performance with the adapter-based method \cite{li2021prefix}. Nevertheless, as we observer results in Fig \ref{fig:fig1} and Table \ref{tab:1}, it exhibits the worst performance on the SV task. This might be caused by the model pre-training phase mostly focusing on the semantics within an utterance. In contrast, the SV task requires the discrimination ability between utterances, which cannot be achieved by modifying the attention weights among input sequences alone. 
For both Bottleneck and MAM adapters, the performance is similar for variants with the same bottleneck dimensionality. However, architectural design renders the MAM adapter more parameter efficient and thus our choice for further experiments. The final metrics improve with the increased dimensionality as shown in Fig \ref{fig:fig1}, where the last data-point for the MAM adapter corresponds to the bottleneck dimensionality of 512, but we set a threshold on the number of parameters to approximately 5M, which in our opinion is a good trade-off between performance and a reasonable model size.
\vspace{-0.5mm}

\subsection{Analysis on in-domain VoxCeleb}
We will first analyze the base scenario with a relatively large amount of in-domain labeled data for fine-tuning. Let us first concentrate on comparing the fine-tuning of pre-trained models (HuBert Base, WavLM Base+, and WavLM Large) via different adapters and the MHFA backend~\cite{peng2022attention}. The bulk of the experiments can be observed in the second and third blocks of the Table~\ref{tab:1}, where we work with the WavLM Base+ and HuBert Base model, respectively. We observe that the MAM adapter performs similarly to the Bottleneck adapter and significantly better than Prefix Tuning across all analyzed Voxceleb test sets and both models. Additionally, we can observe only small degradation when using MAM adapter versus full fine-tuning of all model parameters. We can also safely claim that all PETL methods and full fine-tuning outperform the case when the pre-trained model is fixed and only the MHFA backend is trained. In the third block of Table~\ref{tab:1}, we can also observe the effect of large margin fine-tuning (LM-FT) which consistently improves the performance for both full fine-tuning and MAM adapter strategy across all test conditions. For completeness and consistency with our previous work~\cite{peng2022attention}, we also provide the results with WavLM Large and the MAM adapter including large margin fine-tuning in the last block of Table~ \ref{tab:1}. We can observe a small degradation in performance as now, the total decrease of the learnable parameters is much larger. 

Finally, in the first block of the same table, we can compare our attained results with different approaches selected from the literature. The ECAPA-TDNN represents a standard approach of embedding extractor fully trained from scratch on a supervised dataset (Voxceleb) while the wav2vec-TDNN and UnispeechSAR\_BASE-TDNN represent a combination of a pre-trained model with a TDNN~\cite{snyder2018x} and ECAPA-TDNN~\cite{desplanques2020ecapa} structure for embedding extraction, respectively.

\subsection{Low-resource scenario}
This section presents a scenario where we fine-tune with a small amount of labeled data that we would consider out-of-domain w.r.t. the substantially larger labeled dataset that we call \emph{intermediate dataset} (Voxceleb2-dev). As out-of-domain test set, we chose the CNCeleb-E benchmark and corresponding out-of-domain training sets formed by CNCeleb1 and CNCeleb.T. We perform our experiments with the WavLM Base+ pre-trained model.

Our results are presented in the second block of Table~\ref{tab:3} where we also analyze the impact of the amount of available training data for fine-tuning (adaptation). When confronted with a small amount of training data (200, 400 and 800 speakers), we can observe that direct full fine-tuning (FT) on this data yields the worst results. Only after fine-tuning on CNCeleb.T with 2800 speakers, the performance of direct fine-tuning falls into the same ballpark as other approaches that we will analyze next.

The rather average performance of full fine-tuning and its abrupt degradation with decreasing size of training data would suggest that it is indeed problematic to re-train such a large amount of parameters (94.7M + 2.2M for backend) in a low-resource scenario. The immediate solution might be to leave the pre-trained model fixed and only train the proposed MAM adapter (5.4M + 2.2M parameters). Results with this approach are in the third row of the second block in Table~\ref{tab:3} and we can indeed observe an improvement for low-resource scenarios, but it diminishes when using larger training data such as CNCeleb.T.

In the next two approaches we make use of an intermediate dataset that represents a valuable resource for focusing the large model on the SV task. First, we take the model that is fully fine-tuned on Voxceleb2-dev dataset (first row in the third block of Table~\ref{tab:SOTA}) and further fine-tune it on CNCeleb data. This system is denoted by FT [Int. D] in Table~\ref{tab:3}). We can observe a significant improvement w.r.t. direct fine-tuning and even the direct use of MAM adapter, especially in low-resource scenarios. Again, the improvements diminish with the larger amount of available training data (CNCeleb2). Finally, we start again with the same model, add the MAM adapter and train on CNCeleb data. This yields overall the best results across all analyzed scenarios and even significantly outperforms the previous approaches when larger amount of training data is available. This final approach is especially practical in a sense that we need to store only the parameters of the MAM adapter and the MHFA backend (approximately 5\% of original model size) in order to switch to a new domain while retaining the best possible performance. 



\section{Conclusion}
In this paper, we demonstrate the effectiveness of several PETL methods in the field of speaker verification. The large pre-trained model is frozen, and we only update the inserted lightweight modules. We show that the PETL strategy with MAM adapter is better than simple direct fine-tuning in a low-resource scenario. Additionally, we have demonstrated that having a large labeled intermediate dataset can further improve the overall performance as it preconditions the large transformer-based model for the use in intended task which in our case was speaker verification. Using model directly fine-tuned on such dataset and subsequently training the MAM adapter on low-resource, out-of-domain data, we achieve the best possible performance with the practicality of storing variants of the model for many different domains.


{
\ninept
\bibliographystyle{IEEEbib}
\bibliography{refs}
}
\end{document}